# Big-Five, MPTI, Eysenck or HEXACO: The Ideal Personality Model for Personality-aware Recommendation Systems

Sahraoui Dhelim[1], Liming Luke Chen[2], Nyothiri Aung[1], Wenyin Zhang[3] and Huansheng Ning[1]

[1]School of Computer and Communication Engineering, University of Science and Technology Beijing, China

[2]School of Computing, Ulster University, United Kingdom

[3]School of Information Science and Engineering, Linyi University, China

**Abstract**

Personality-aware recommendation systems have been proven to achieve high accuracy compared to conventional recommendation systems. In addition to that, personality-aware recommendation systems could help alleviate cold start and data sparsity problems. Most of the existing works use Big-Five personality model to represent the user's personality, this is due to the popularity of Big-Five model in the literature of psychology. However, from personality computing perspective, the choice of the most suitable personality model that satisfy the requirements of the recommendation application and the recommended content type still needs further investigation. In this paper, we study and compare four personality-aware recommendation systems based on different personality models, namely Big-Five, Eysenck and HEXACO from the personality traits theory, and Myers–Briggs Type Indicator (MPTI) from the personality types theory. Following that, we propose a hybrid personality model for recommendation that takes advantage of the personality traits models, as well as the personality types models. Through extensive experiments on recommendation dataset, we prove the efficiency of the proposed model, especially in cold start settings.

## Introduction

Personality Computing has emerged as a new study field that aims to capture, manipulate, and make use of the human personality character through the use of information and communication technologies. Personality computing can be view as the emerging domain that comes as a result of the decoupling of information technologies and psychology personality theory, as shown in Figure 1. While most of the previous works in the field of personality computing have focused on Automatic Personality Recognition (APR) by analyzing the user's data [1], and the use of personality traits to empower robots to become more social during Human-Robot interaction [2]. Another important application of personality computing is personality-aware recommendation systems [3]. As the conventional recommendation systems suffer from many problems, such as cold start problem where there is not enough data to infer the user's preferences, data sparsity and free riders, to name a few. Personality-aware recommendations was proposed as a new method that can achieve high accuracy compared to the conventional recommendation systems and alleviate the effects of cold start and data sparsity problems. However, most of the existing works uses Big-Five personality model to represent the user's personality (also known as Five Factors Model), this is due to the popularity of Big-Five model in the literature of psychology. However, from a personality computing perspective, the choice of the most suitable personality model that satisfies the requirements of the recommendation application and the recommended content type still needs further investigation. In this paper, we study and compare four personality-aware recommendation systems based on different personality models, namely Big-Five, Eysenck and HEXACO from the personality traits theory, and Myers–Briggs Type Indicator (MPTI) from the personality types theory. Following that, we propose hybrid personality-aware recommendation system that take advantage of the personality traits models, as well as the personality types models. Through extensive experiments on recommendation dataset, we prove the efficiency of the proposed model, especially in cold start settings.

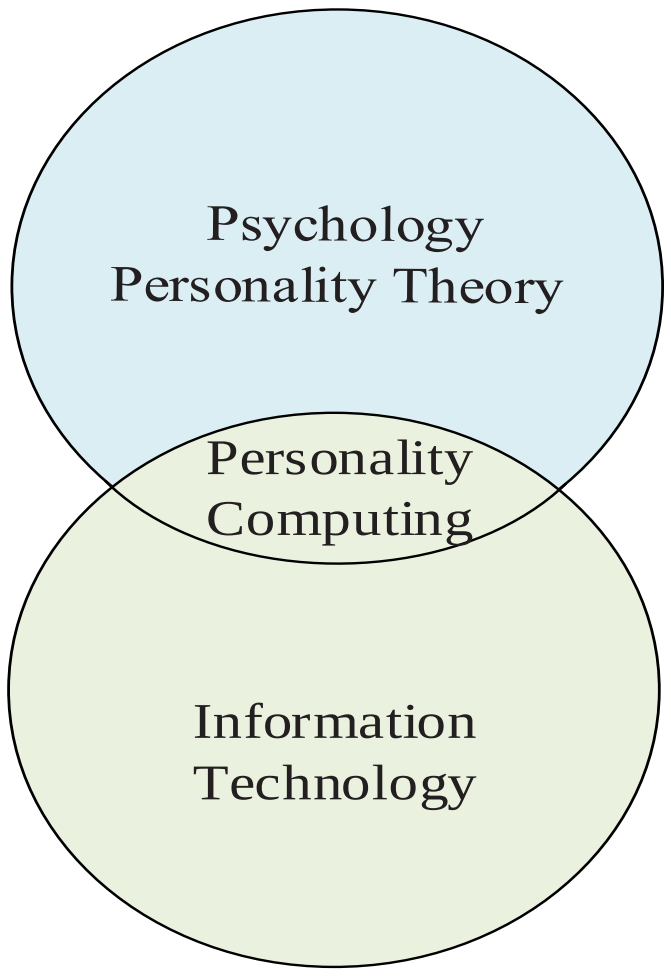

**Figure 1** Personality computing scope

The contributions of this paper is summarized as follows

- Design and compare four personality-aware recommendation systems based on Big-Five, Eysenck, HEXACO and MPTI personality models
- Propose a hybrid personality-aware recommendation system that take advantage of the personality traits models, as well as the personality types models.
- Perform comparative experiment by applying the proposed personality-aware recommendation systems on news recommendation dataset.

The rest of the paper is organized as follows:

In Section 2, we review the related works that have used the three personality models in the context of recommendation systems. While in Section 3, we introduce the four personality models, as well as their measurement methods. In Section 4, we introduce the fundamental concept of personality-aware recommendation and how personality traits could be incorporated in the recommendation process. In Section 5, we show the details of the conducted experiment and the evaluation process and discuss the obtained results, and finally, the paper is concluded in Section 7.

## 2.Related works

Personality traits have been used to enhance the recommendation systems in many domains, and many researchers have advocated for the massive adaptation of the user's personality characteristics and other social features in recommendation systems. Vinciarelli et al. [4] surveyed the field of personality computing and applications of user personality in computing systems. Similarly, Kaushal et al. [5] surveyed the recent automatic personality recognition schemes and presented the trends of each recognition technique. Dhelim et al. [3] have surveyed the literature of personality-aware recommendation systems. Ning et al. [6]

proposed a personality-aware friend recommendation system named PersoNet that leverages Big-Five personality traits to enhance the hybrid filtering friend selection process. PersoNet outperformed the conventional rating-based hybrid filtering, and achieve an acceptable precision and recall values in cold start phase as well. Similarly, Chakrabarty et al. [7] designed a personality-aware friend recommendation system name FAFinder (Friend Affinity Finder). FAFinder uses Hellinger-Bhattacharyya Distance (H-B Distance) to measure the user's Big-Five similarity and recommend friends accordingly. Dhelim et al. [8] used Big-Five personality traits to improve the accuracy of user interest mining. Besides recommendation systems, users' personality was incorporated as a social feature that could be used to understand the social context of the users. While in [9] propose Meta-Interest, a personality-aware product recommendation system based on user interest mining and meta path discovery. Their proposed system detects the user's interest and the items associated with these interests to perform the recommendations. In other work [10], the authors discussed the usage of personality information in the context of smart home scenario, and in [11] the authors discussed the excessive usage of technology on psychological disorders and its effect on the user's personality. In [12] the user's personality was represented as a thinking entity that is represented by a cyber entity in the cyberspace. A personality-aware recommendation systems were proposed in [9], where the products were associated with the users' topical interests and personality traits. Similarly, in [13] a user interest mining scheme leverages the user's personality traits in the context of social signed networks. Many works have used personality traits for academic-oriented recommendation systems, such as courses recommendations, conference attendee recommendations and research paper recommendations. Xie el al [14] proposed a recommendation system of academic conference participants called SPARP (Socially-Personality-Aware-Recommendation-of-Participants). For more effective collaborations in the vision of a smart conference, the proposed recommendation approach uses a hybrid model of interpersonal relationships among academic conference participants and their personality traits. At first, the proposed system determines the social ties among the participants based past and present social ties from the dataset with four trial-weight parameters. These weight parameters are used later in their experiment to represent various influence proportions of the past and present social ties among participants. Following that, the system calculates the personality-similarity between the conference participants based on explicit tagged-data of the personality ratings. Fahim Uddin et al [15] Proposed a personality-aware framework to improve academic choice for new enrolled students.Their proposed framework makes use of the research field of Predicting Educational Relevance For an Efficient Classification of Talent, that uses stochastic probability distribution modeling to help student to choose the relevance academic field. Hariadi et el [16] proposed a personality-aware book recommendation system that combine the user's attributes as well as his personality traits. The proposed system leverages MSV-MSL (Most Similar Visited Material to the Most Similar Learner) method to compute the similarity between users and form the personality neighborhood. Hill et al [17] investigated the association between HEXACO personality model with preferences for certain aspects of gaming experiences. The main finding confirmed that extraversion trait is moderately associated with the socializer gaming preference and a slight association with the daredevil gaming preference.

## 3.Personality models

Since the early ages of Greek philosophers, scientists have agreed on the importance of the personality study as a vital factor to understand individual behaviors and way of thinking. There is no universal theory that explains the human personality comprehensively. Some theories explain the difference in personality to genetics, while others associate it with sociological factors. There are many personality models that have been extensively studied from psychological perspective such as Big-Five personality traits model, MBTI, Eysenck personality model and HEXACO personality model. These personality models differ in the way they represent the human personality, some assume that the human has “types” of personality (MBTI), while others represent the personality as a spectral of personality traits (Big-Five, Eysenck and HEXACO). From a personality computing perspective the personality traits theory such as Big-Five model have been applied in most of the previous personality computing works.

### 3.1.Big-Five model

The Big-Five personality traits model [18], also famous as five-factor model (FFM) is the most used model in psychology as well as personality computing works. The Big Five model mainly defines the five factors as Openness to experience, Conscientiousness, Extraversion, Agreeableness and Neuroticism. And often these traits are represented by the acronyms OCEAN or CANOE, as shown in Figure 2. Some of the related characters also known as facets of the Big Five personality traits are listed in Table I.

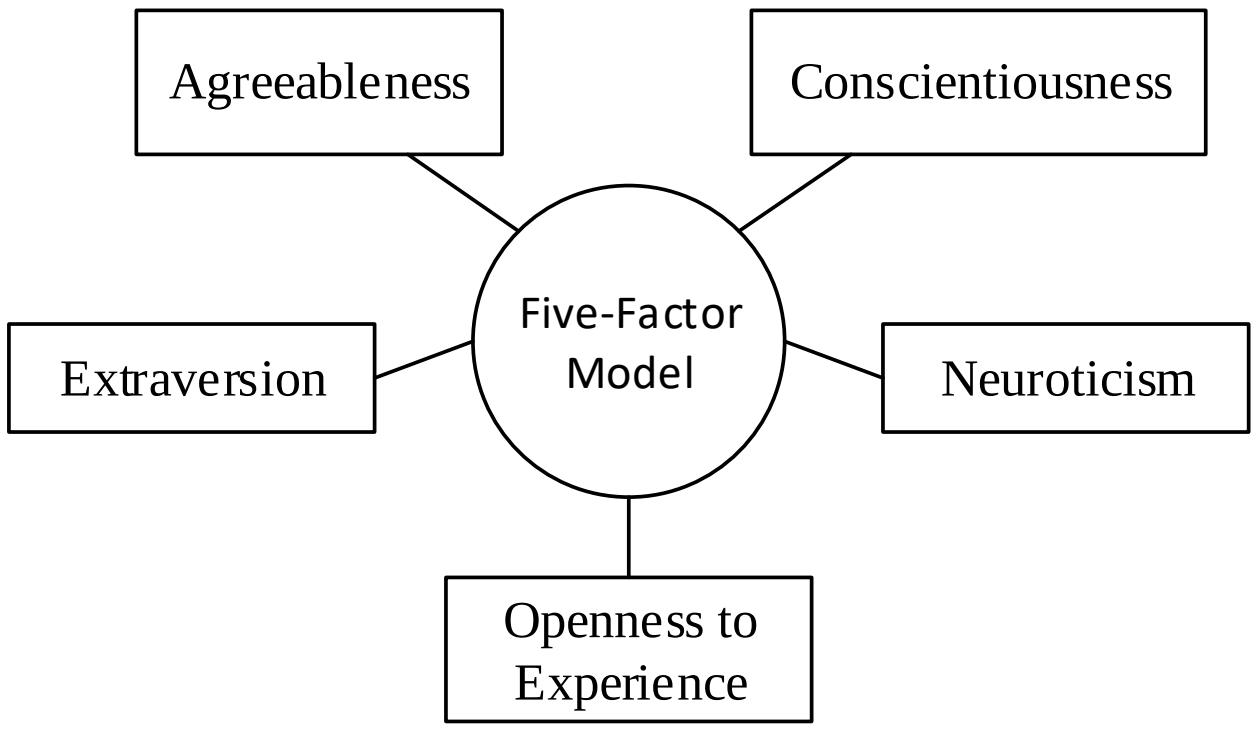

**Figure 2** Big Five personality traits

**Table I** Big Five Traits and Associated Characters

| Personality Trait | Related Characters |
|---|---|
| Openness to Experience | Insightful, Curious, Wide interests, Imaginative, Artistic, Original |
| Agreeableness | Trusting, Appreciative, Kind, Forgiving, Generous, Sympathetic |
| Conscientiousness | Reliable, Efficient, Planful, Responsible, Thorough, Organized |
| Extraversion | Energetic, Active, Assertive, Outgoing, Talkative |
| Neuroticism | Tense, Anxious, Touchy, Worrying, Unstable, Self-pitying |

### 3.2.Eysenck personality model

As its name indicates, this theory was proposed by Hans Eysenck [19]. Eysenck theory is mainly based on genetics and physiology, however he also stressed that personality could also be shaped by sociological factors. Eysenck theory assumes that the human personality could be identified by measuring three independent dimensions of temperament, mainly Extraversion/Introversion (E), Neuroticism/Stability (N) and Psychoticism/Socialisation (P) as shown in Figure 3. The characters related to each personality dimension are listed in Table II

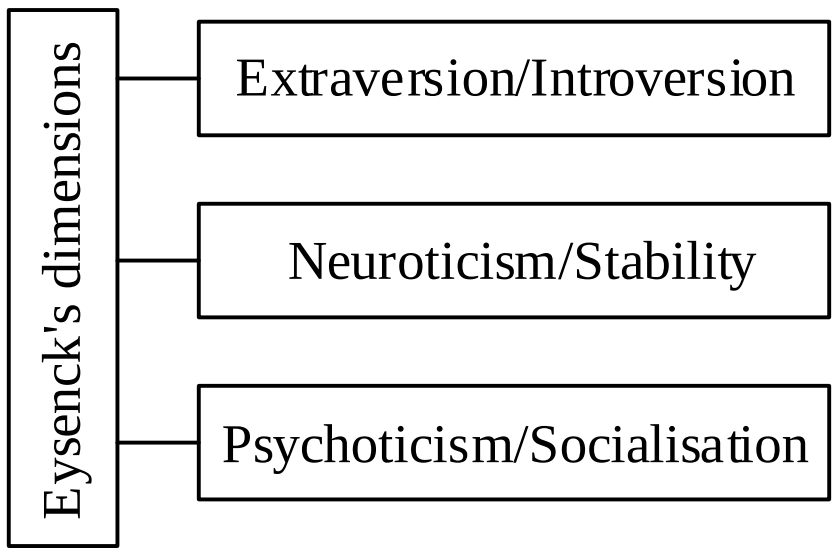

**Figure 3** Eysenck personality dimensions

**Table II** Eysenck dimensions and Associated Characters

| **Psychoticism** | **Extraversion** | **Neuroticism** |
|---|---|---|
| Aggressive | Sociable | Anxious |
| Assertive | Irresponsible | Depressed |
| Egocentric | Dominant | Guilt Feelings |
| Unsympathetic | Lack of reflection | Low self-esteem |
| Manipulative | Sensation-seeking | Tense |
| Achievement-oriented | Impulsive | Moody |
| Dogmatic | Risk-taking | Hypochondriac |
| Masculine | Expressive | Lack of autonomy |
| Tough-minded | Active | Obsessive |

### 3.3.HEXACO personality model

The HEXACO personality model is an extension of the Big-Five model. However HEXACO model add a new dimension known as the Honesty-Humility dimension to the other five personality traits of Big-Five model. HEXACO was proposed by Ashton and Lee [20]. Figure 4 shows the dimensions of HEXACO model and the characters associated with each dimension are presented in Table III.

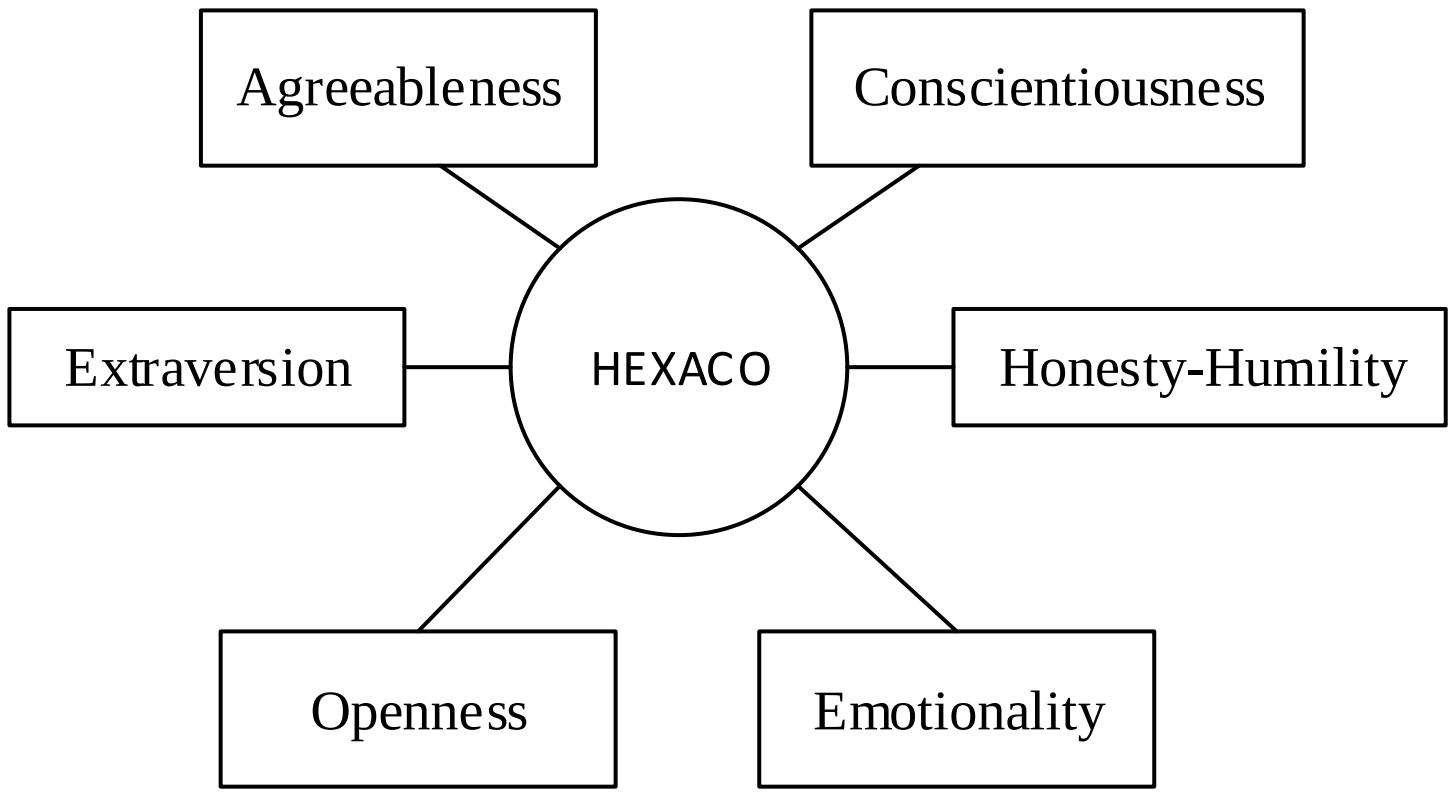

**Figure 4** HEXACO personality traits

**Table III** Eysenck dimensions and Associated Characters

| **HEXACO dimension** | **Associate characters** |
|---|---|
| Honesty-Humility | Sincerity, Fairness, Greed Avoidance, Modesty |
| Emotionality | Fearfulness, Anxiety, Dependence, Sentimentality |
| Extraversion | Social Self-Esteem, Social Boldness, Sociability, Liveliness |
| Agreeableness | Forgivingness, Gentleness, Flexibility, Patience |
| Conscientiousness | Organization, Diligence, Perfectionism, Prudence |
| Openness | Aesthetic Appreciation, Inquisitiveness, Creativity, Unconventionality |

### 3.4. MPTI

Another personality model rarely used in personality computing is the Myers Briggs Type Indicator (MBTI) [21], unlike Big-Five and HEXACO, MBTI defines the personality as types rather than traits, in other words, the human personality is exclusively defined by one personality type/class, rather than having a different score in multiple traits. MBTI defines four categories: introversion or extraversion, sensing or intuition, thinking or feeling, judging or perceiving. One letter from each category is taken to produce a four-letter personality types, which makes 16 possible personality types: ISTJ, ISFJ, INFJ, INTJ, ISTP, ISFP, INFP, INTP, ESTP, ESFP, ENFP, ENTP, ESTJ, ESFJ, ENFJ and ENTJ.

### 3.5. Personality measurement

There are many methods to measure one’s personality; questionnaires where subjects answer with Likert scale question about how they describe themselves are the most common personality measurement method. There are many well-known personality questionnaires with different sizes (item number). NEO-Personality-Inventory Revised (NEO-PI-R, 240 items) is one of the most adopted long personality questionnaires [22]. For medium size questionnaires, the NEO Five-Factor Inventory (NEO-FFI, 60 items), and the Big Five Inventory (BFI, 44 items) is used frequently [23]. Some other short questionnaires are much faster to fill (5-10 items), such as BFI-10 [24], short questionnaires retain only the most correlated items with each personality trait. Table 2 shows the items of BFI-10 Big-Five questionnaire.

**Table 1 The BFI-10 Personality questionnaire**

| Item | Question | Dimension |
|---|---|---|
| 1 | I am outgoing, sociable | Extraversion |
| 2 | I get nervous easily | Neuroticism |
| 3 | I tend to be lazy | Conscientiousness |
| 4 | I have an active imagination | Openness |
| 5 | I am reserved | Extraversion |
| 6 | I am generally trusting | Agreeableness |
| 7 | I have few artistic interests | Openness |
| 8 | I tend to find fault with others | Agreeableness |
| 9 | I do a thorough job | Conscientiousness |
| 10 | I am relaxed, handle stress well | Neuroticism |

## 4.Personality-aware recommendation

Personality neighborhood filtering is the most common personality-aware recommendation technique. Typically, the system uses a proximity function that measures the personality similarity to find the personality neighborhood users, and use it to predict future rating accordingly. The system design of the proposed personality-aware recommendation systems is illustrated in Figure 1. The first step is the personality measurement, where the system extract the user's personality information, either by asking the user to answer a personality questionnaire or by applying an automatic personality recognition scheme [25]. The second step is the personality similarity measurement, in which the system tries to associate the newly joined user with the most similar neighbors in terms of personality types. This step enables the personality-aware recommendation system to offer recommendation based only on personality information, which alleviates the cold-start problem [26]. When the user starts to give rating and passes the cold-start phase, the recommendation system refines the set of neighbors by incorporating the user rating in the overall similarity measurement.

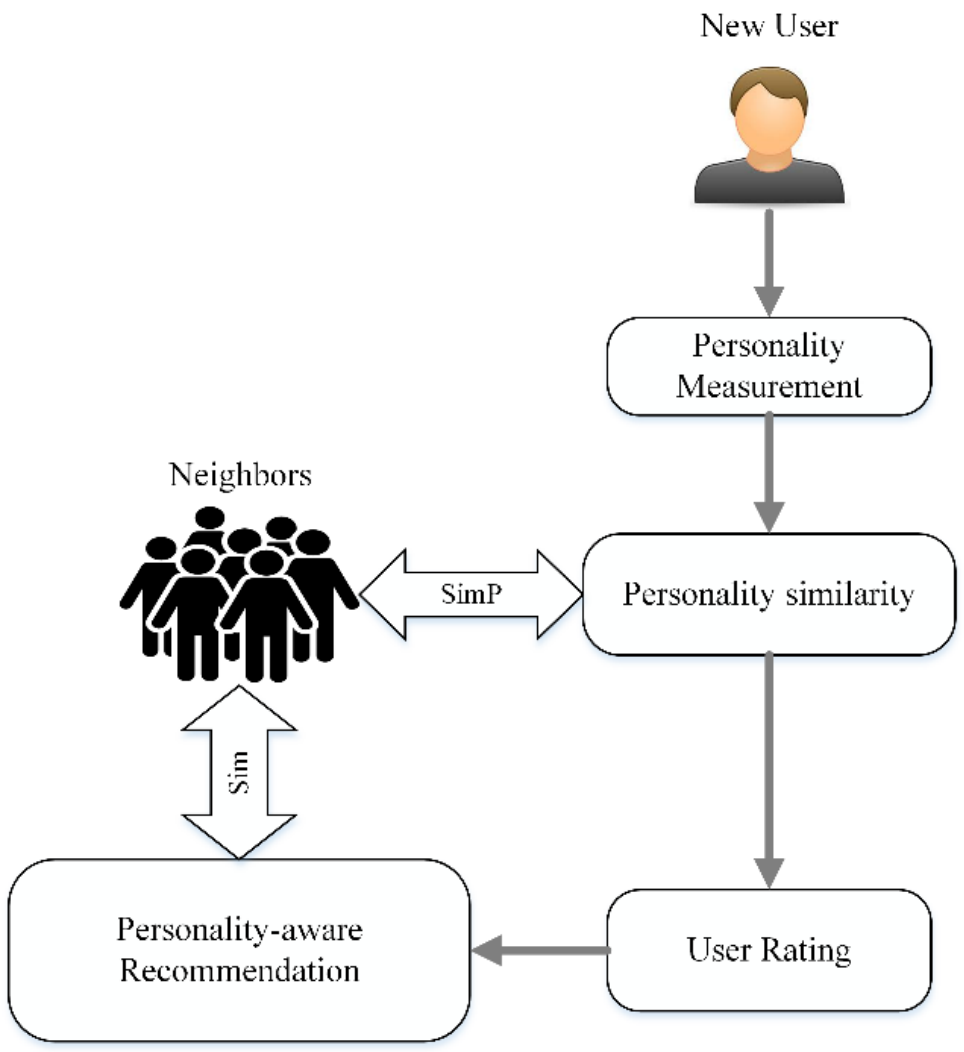

Figure 1 Personality-aware recommendation steps

There are many similarity measurement methods that can be used to measure the proximity between two users, Pearson correlation coefficient is the most commonly used proximity function. Given two users $u_x$ and $u_y$, the rating similarity between them is computed using the function $SimR(u_x, u_y)$ as shown in (1), where $R_x$ and $R_y$ is the sets of previous rating of user $u_x$ and $u_y$ respectively, and $r_{x,i}$ is the rating of user $u_x$ on item $i$, and $\overline{r_x}$ is the mean rating of user $u_x$.

$$SimR(u_x, u_y) = \frac{\sum_{i \in R_x \cap R_y} (r_{x,i} - \overline{r_x})(r_{y,i} - \overline{r_y})}{\sqrt{\sum_{i \in R_x \cap R_y} (r_{x,i} - \overline{r_x})^2 \sum_{i \in R_x \cap R_y} (r_{y,i} - \overline{r_y})^2}} \quad (1)$$

We adopt Pearson correlation coefficient to measure the personality similarity between users, as shown in (2), where $\overline{p_x}$ and $\overline{p_y}$ is the average value of the personality traits vector for user $u_x$ and $u_y$ respectively, and $p_x^i$ is the $i^{th}$ trait in the personality traits vector. To compare the three studied personality models (Big-five, Eysenck and HEXACO), we implement three recommendation systems by changing the personality similarity function SimP to measure the similarity of users using their respective personality model.

$$SimP(u_x, u_y) = \frac{\sum_i (p_x^i - \overline{p_x})(p_y^i - \overline{p_y})}{\sqrt{\sum_i (p_x^i - \overline{p_x})^2 \sum_i (p_y^i - \overline{p_y})^2}} \quad (2)$$

The overall similarity measurement between users $u_x$ and $u_y$ is computed using the function Sim, as shown in (3)

$$Sim(u_x, u_y) = \alpha \times SimP(u_x, u_y) + (1 - \alpha) \times SimR(u_x, u_y) \quad (3)$$

where $\alpha$ is the cold-start parameter that adjusts the contribution of personality-based similarity to the overall similarity measurement $(1 \geq \alpha \geq 0)$, and it is negatively correlated with the number of neighbors. After computing the similarity among users and eventually establishing the personality neighborhood of each user, the prediction score is computed by aggregating the rating of neighborhood users and the similarity with these users. Formally, let $score(u, i)$ denote the predication score that user u will give to item i, the prediction score is computed as shown in (4).

$$score(u, i) = \overline{r_u} + k \sum_{v \in \Omega_u} sim(u, v)\ (r_{v,i} - \overline{r_v}) \quad (4)$$

where $\overline{r_u}$ and $\overline{r_v}$ are the average rating of user $u$ and user $v$ respectively, and $r_{v,i}$ is the rating given by user $v$ to item $i$, and $\Omega_u$ are the neighbors of user $u$ that have previously rated item $i$. The total similarity $sim(u, v)$ is the product of the rating similarity and personality similarity.

The above-mentioned personality-aware recommendation model is applied to compare the four personality models. For our proposed hybrid personality-aware recommendation system that combines the personality traits theory and the personality type theory, we extend the models as shown in Algorithm 1, where $\lambda$ is the personality similarity threshold and $\delta$ is the overall similarity threshold, while MPTI($u_x$) is a function that return the MPTI personality type of user $u_x$, and $N_x$ is the set of neighbors.

| Algorithm 1: Hybrid_Personality_Recommender($u_x$) |
|---|
| IF(COLDSTART) THEN<br>FOREACH $u_y \in U$ Do<br>IF ($SimP(u_x, u_y) > \lambda$) OR (MPTI($u_x$)=MPTI($u_y$))THEN<br>$N_x \leftarrow N_x \cup \{u_y\}$<br>ENDIF<br>ENDFOR<br>ELSE<br>FOREACH $u_y \in U$ Do<br>IF ($Sim(u_x, u_y) > \delta$) AND (MPTI($u_x$)=MPTI($u_y$))THEN<br>$N_x \leftarrow N_x \cup \{u_y\}$<br>ENDIF<br>ENDFOR<br>ENDIF |

## 5.Experiment and evaluation

To compare the three personality model (Big-Five, Eysenck and HEXACO), we have implemented three personality-aware recommendation systems based on these personality models.

### 5.1.Evaluation dataset

The three systems were evaluated using Newsfullness news datasets [8]. The dataset contains the personality information of users that was obtained during the users' registration, and the articles viewing history of each user along with the labels related to each article. After the preprocessing step, in which we remove the users that have very few viewed articles, we used the data of 1229 users, who have viewed 33450 articles for the period of 3 months.

### 5.2.Evaluation metrics

After computing the personality similarity using the four personality model (Big-Five, Eysenck, HEXACO and MPTI), each personality-aware recommendation system computes the set of neighbors and recommend the relevant items accordingly. The four personality-aware recommendation systems were tested based on their precision that measures the ability of the recommendation system to compute all the relevant articles, recall that measure the ability of the correctness of the recommended items and f-measure as a measure that represents the combination of precision and recall. Specifically, we use the three studied personality-aware recommendation systems to calculated the articles that are relevant to each user. Formally, Let $F = R \cup I$ be the set of all articles that were displayed to user $u$, where $R = \{z_1, z_2, \ldots, z_r\}$ is the set relevant articles, and $I = \{z_1, z_2, \ldots, z_i\}$ is the set of irrelevant articles. Let $V = \{z_1, z_2, \ldots, z_v\}$ be the set of viewed articles. At this point, we want to measure the following values: (1) true positives: the set of relevant articles that the user has viewed $TP = \{x \mid x \in R \cap V\}$, (2) false positives: the set of irrelevant articles that viewed by the user $FP = \{x \mid x \in I \cap V\}$ and (3) false negatives: the set of relevant articles that not viewed by the user $FN = \{x \mid x \in R, x \notin V\}$. Based on that we computed the following evaluation metrics:

Precision: the portion of relevant viewed articles in the total viewed articles, and it is computing using (5)

$$Precision = \frac{TP}{TP+FP} \quad (5)$$

Recall: the portion of relevant viewed articles in the total relevant articles, and it is computing using (6)

$$Recall = \frac{TP}{TP+FN} \qquad (6)$$

F-Measure: also called the balanced F-Score, it is the harmonic average of the precision and recall, and it is computing using (7)

$$F = \frac{2\,P\,R}{P+R} \qquad (7)$$

**5.3.Results analysis**

The user classification according to the users' dominant personality traits according to Big-five, HEXACO and Eysenck models are presented in Figure 2, Figure 3 and Figure 4 respectively. As we can observe the extraversion trait is the most dominant trait in the three models. While neuroticism is the least common traits among all users for all the personality models.

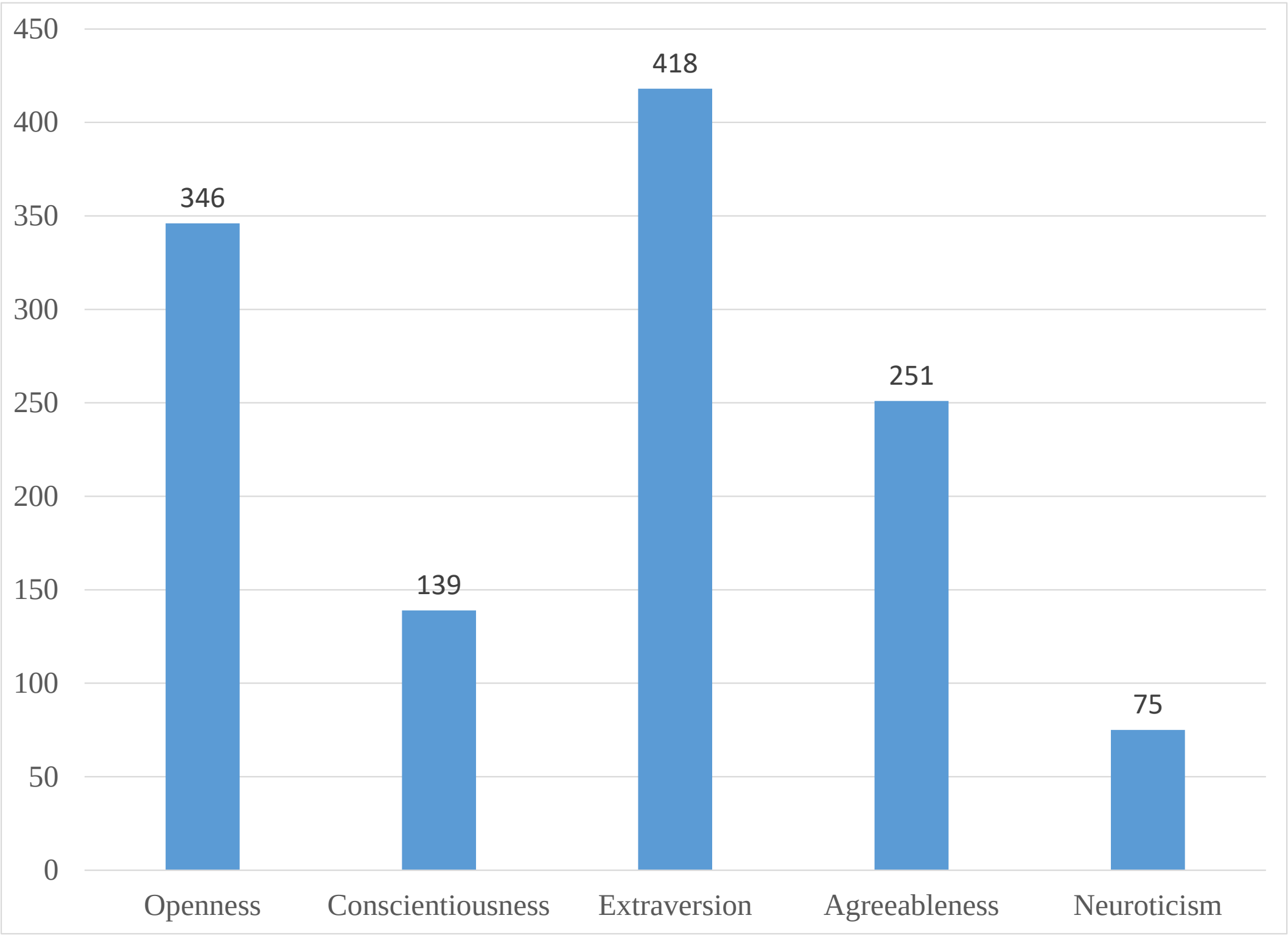

Figure 2 Big-five users classification

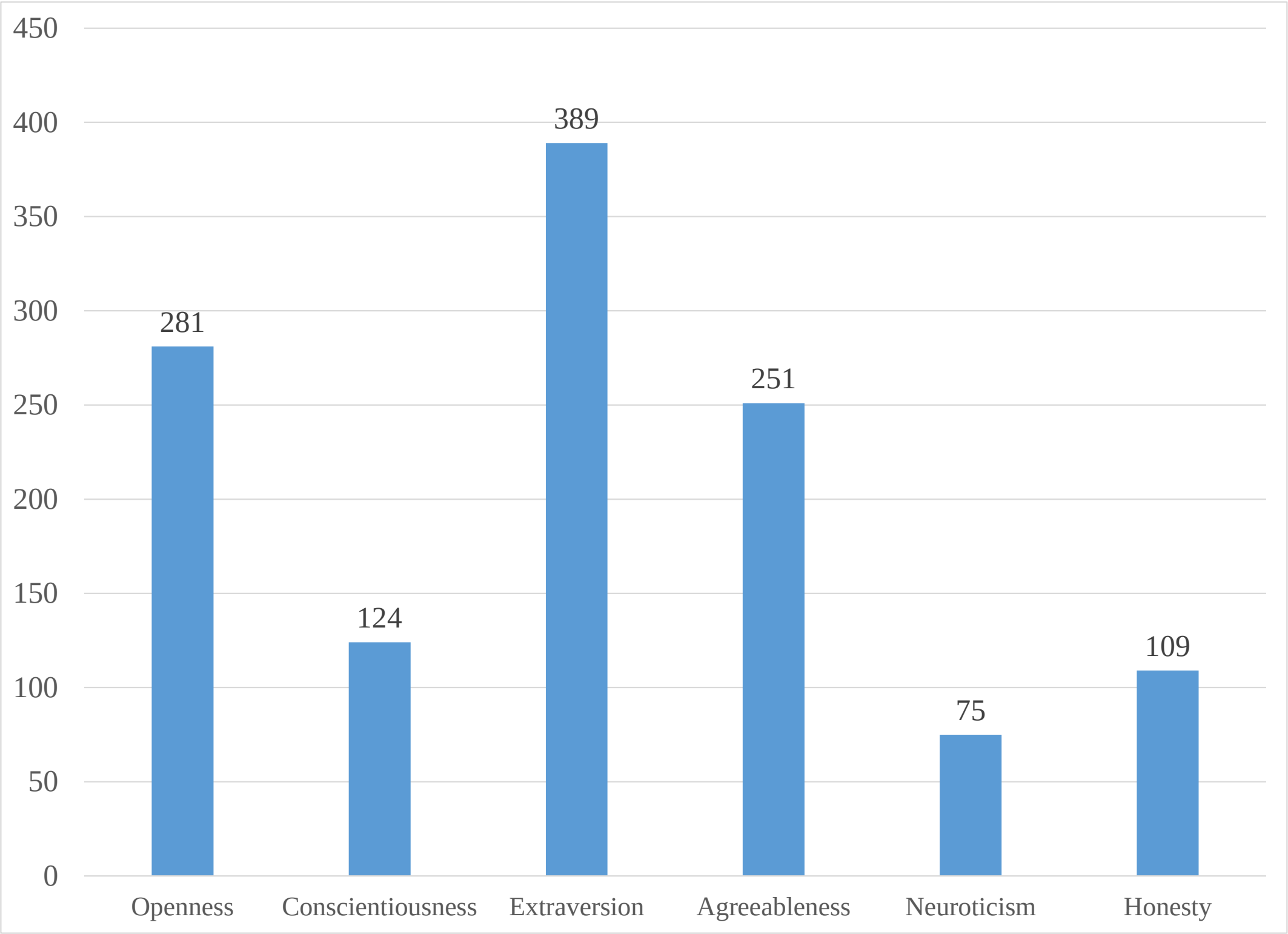

Figure 3 HEXACO users classification

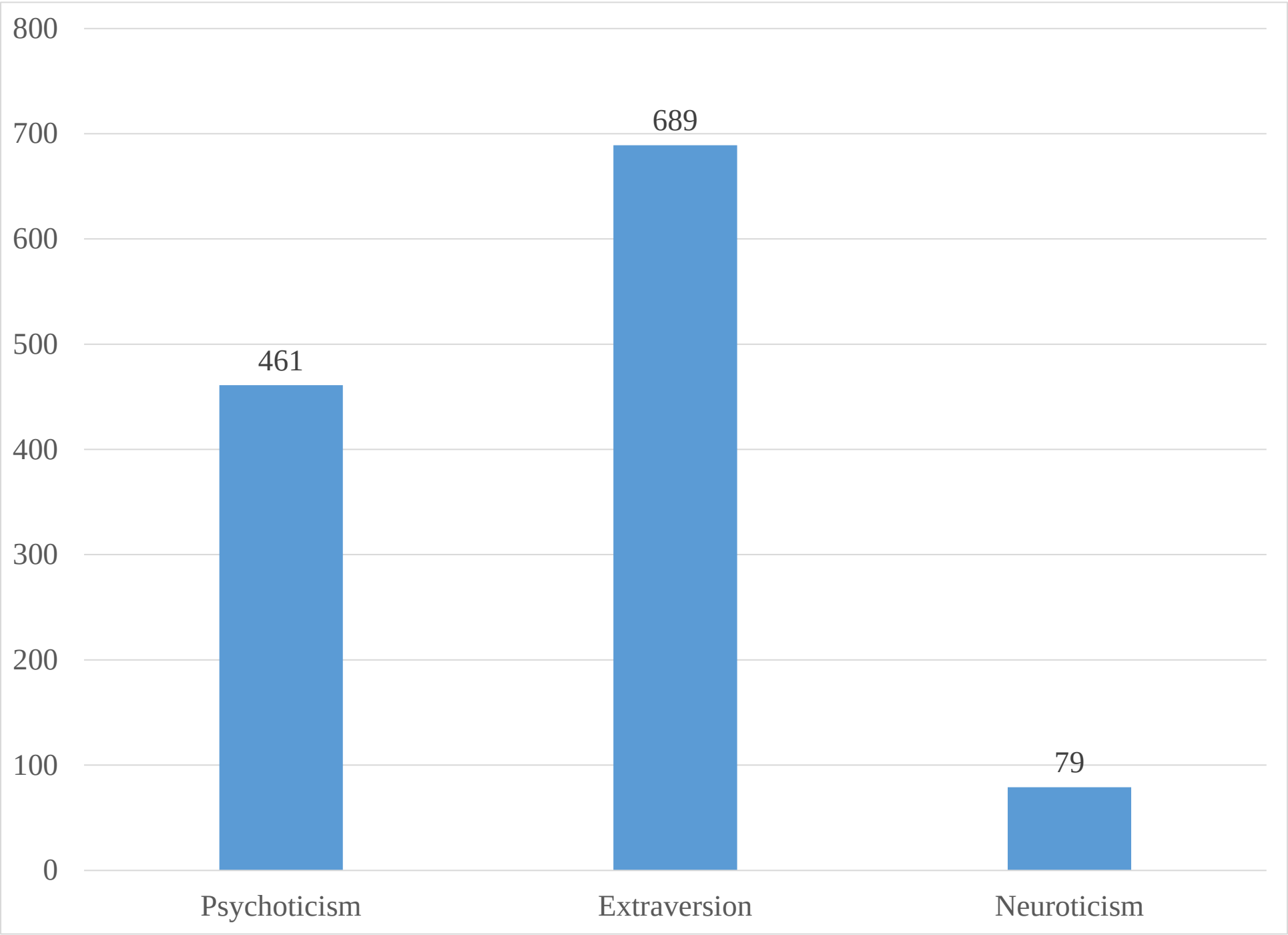

Figure 4 Eysenck users classification

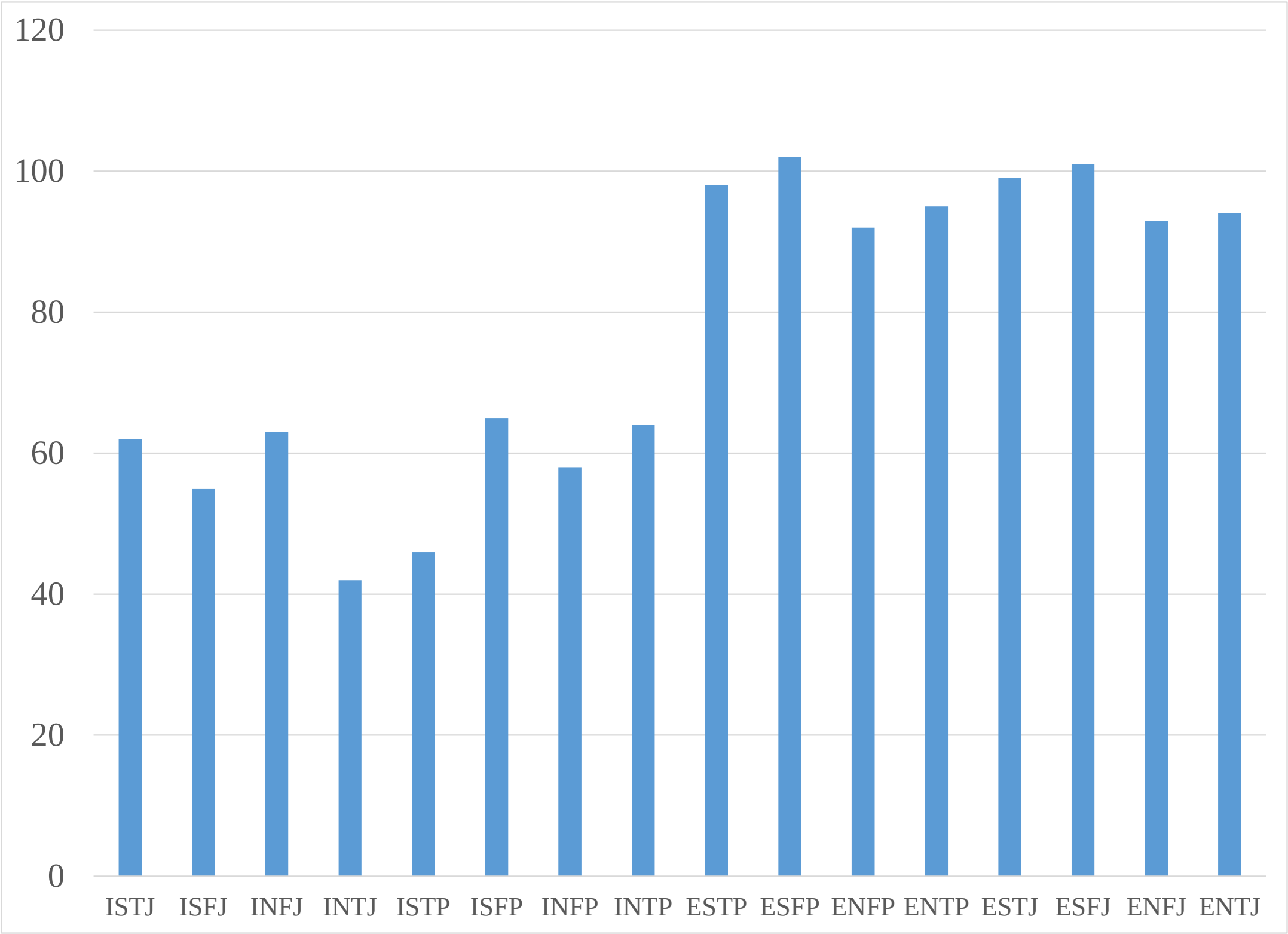

Figure 5 MPTI users classification

The performance of the three personality-aware recommendation systems in terms of precision, recall and F-Measure is presented in Figure 6, Figure 7 and Figure 8 respectively. Figure 6 represents the average precision value when different values for the previously view articles count, we can observe that Eysenck model has a better precision value with few viewed articles count in the cold start phase compared to other personality traits models (Big-Five and HEXACO). That is because Eysenck model has only three traits, which makes categorizing users more generic. We can also notice that MPTI also performs better than (Big-Five and HEXACO), that is because MPTI is a personality type theory rather than personality trait theory, therefore it is relatively easier to find similar users with the same personality type than computing similarity with a spectrum of traits. However, when the users pass the cold start phase and view enough articles, the similarity computed with personality traits (Big-Five and HEXACO) are more accurate in computing similarities among users. From Figure 6, we can also observe that our proposed hybrid personality model has the best performance, that is because it leverages the advantages of personality type model at the cold start phase, and also the advantages of personality traits theory at later stages. Similarly, Figure 7 shows that Hybrid, Eysenck and MPTI also have a better recall in the cold start phase, and personality traits (Big-Five and HEXACO) have the upper hand ultimately. From Figure 6 and Figure 7 we can also notice that HEXACO slightly outperforms Big-Five due to the additional sixth trait (Honesty-Humility). Figure 8 shows the F-Measure values that combine both the precision and recall in one figure.

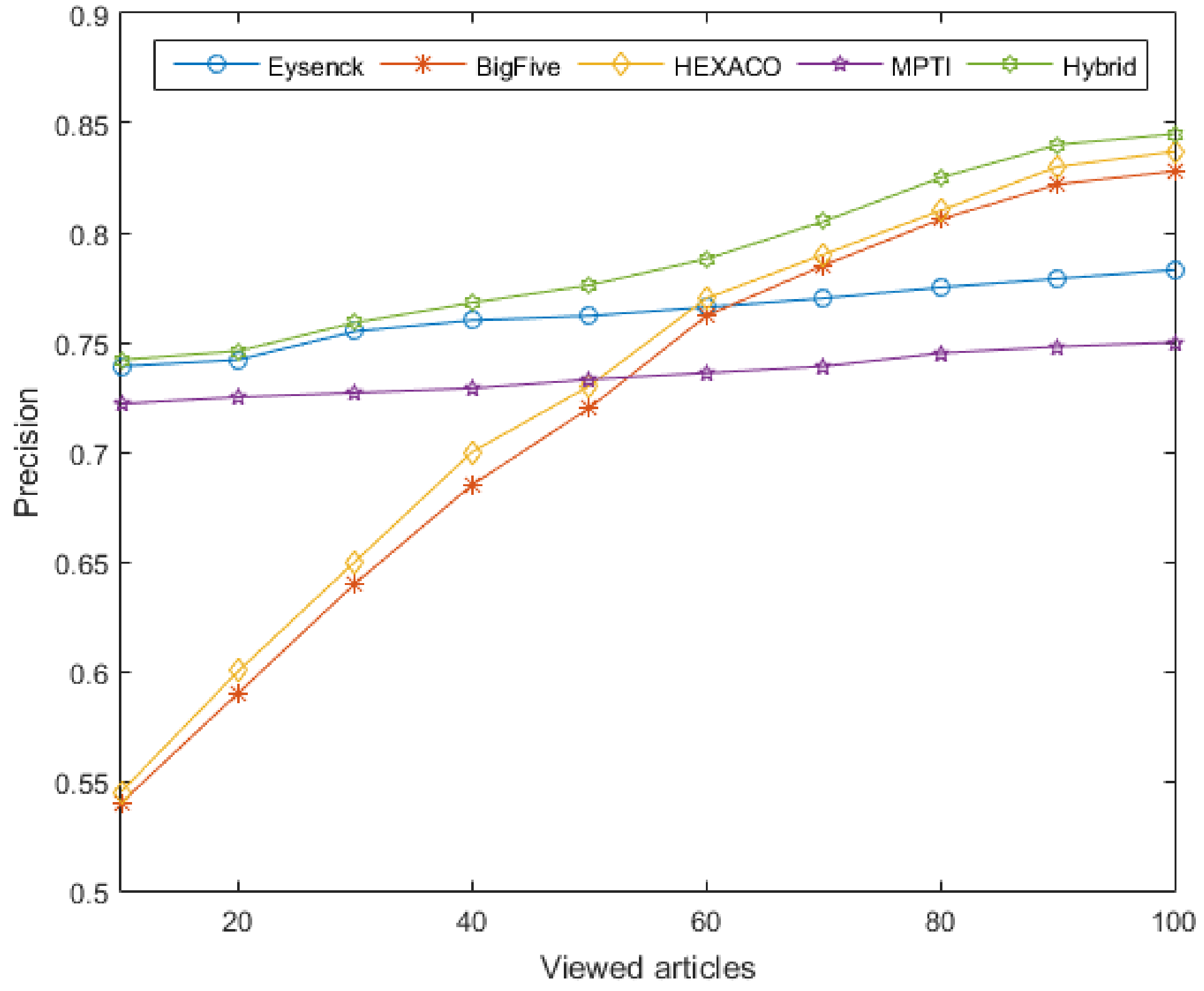

Figure 6 Precision vs article count

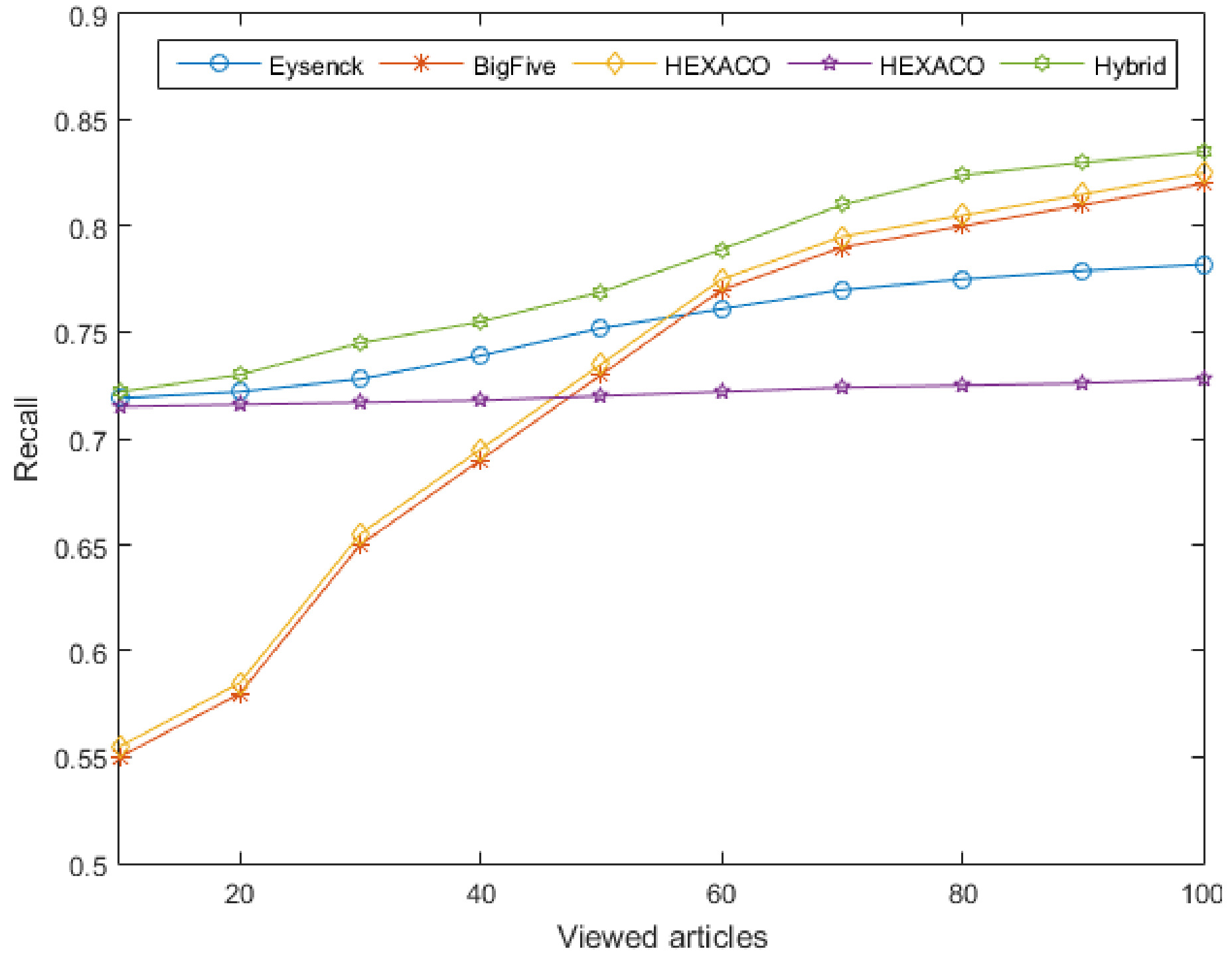

Figure 7 Recall vs article count

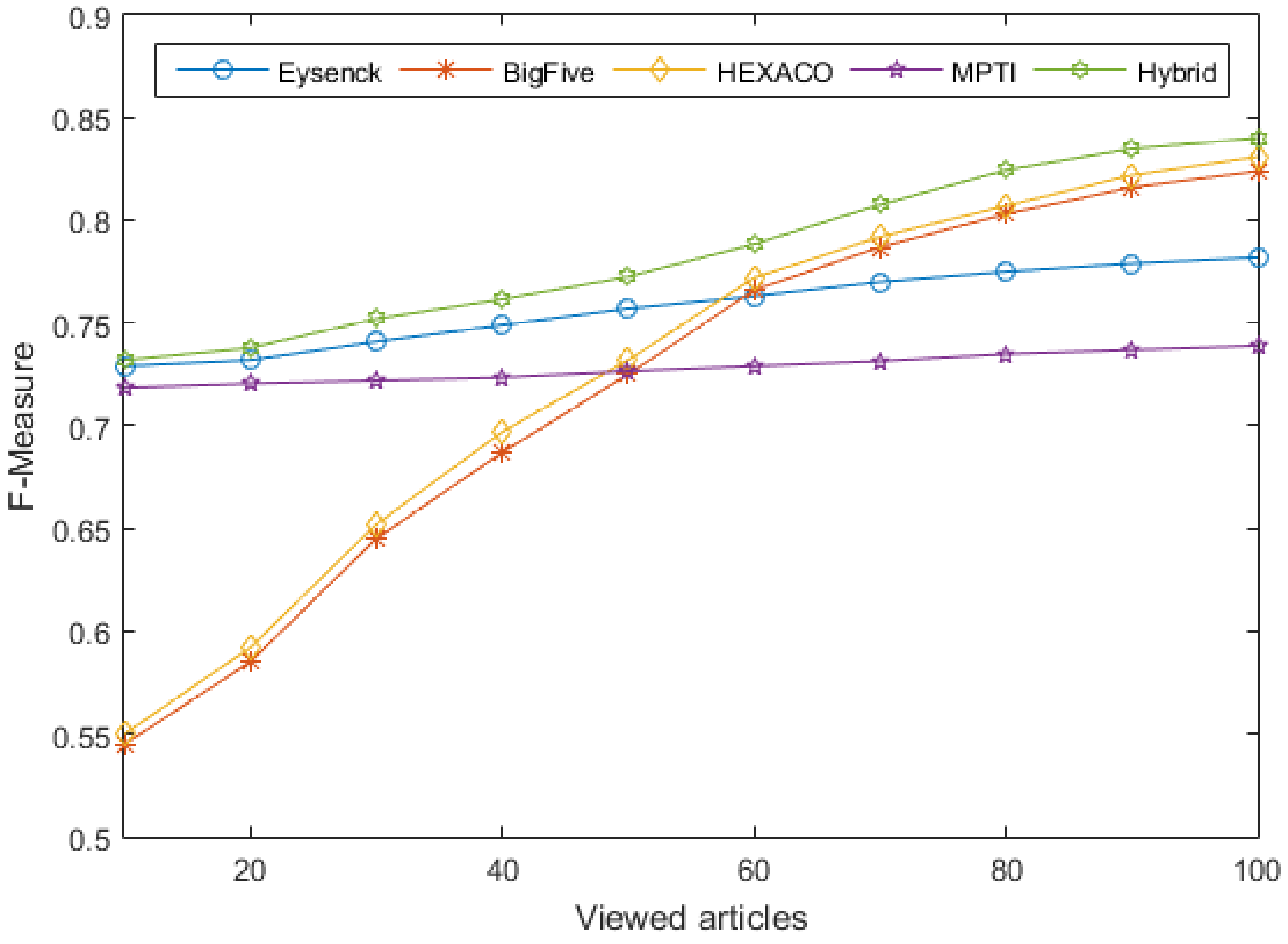

Figure 8 F-Measure vs article count

## Conclusions

In this paper, we have studied and compared four personality models (Big-Five, Eysenck, MPTI and HEXACO) in the context of recommendation systems. Moreover, we have proposed a new hybrid personality model for recommendation systems that takes advantage of the personality traits models, as well as the personality types models. The obtained results confirm that our proposed model is well suited for personality-aware recommendation systems, as it leverages the personality type model to mitigate the cold start problem, and also incorporates the advantages of the personality traits model. Eysenck model is well suited to alleviate the cold start effects more than the personality traits models (Big-Five and HEXACO). However, when the users pass the cold start phase and view enough articles, the similarity computed with personality traits (Big-Five and HEXACO) are more accurate in computing similarities among users. The results also show that HEXACO slightly outperforms Big-Five due to the additional sixth trait (Honesty-Humility).

## References

[1] N. Majumder, S. Poria, A. Gelbukh, and E. Cambria, “Deep Learning-Based Document Modeling for Personality Detection from Text,” *IEEE Intell. Syst.*, vol. 32, no. 2, pp. 74–79, Mar. 2017.

[2] B. Tay, Y. Jung, and T. Park, “When stereotypes meet robots: the double-edge sword of robot gender and personality in human--robot interaction,” *Comput. Human Behav.*, vol. 38, pp. 75–84, 2014.

[3] S. Dhelim, N. Aung, M. A. Bouras, H. Ning, and E. Cambria, “A Survey on Personality-Aware

Recommendation Systems," Jan. 2021.

[4] A. Vinciarelli and G. Mohammadi, "A survey of personality computing," *IEEE Trans. Affect. Comput.*, vol. 5, no. 3, pp. 273–291, 2014.

[5] V. Kaushal and M. Patwardhan, "Emerging Trends in Personality Identification Using Online Social Networks—A Literature Survey," *ACM Trans. Knowl. Discov. from Data*, vol. 12, no. 2, p. 15, 2018.

[6] H. Ning, S. Dhelim, and N. Aung, "PersoNet: Friend Recommendation System Based on Big-Five Personality Traits and Hybrid Filtering," *IEEE Trans. Comput. Soc. Syst.*, pp. 1–9, 2019.

[7] N. Chakrabarty, S. Chowdhury, S. D. Kanni, and S. Mukherjee, "FAFinder: Friend Suggestion System for Social Networking," in *Intelligent Data Communication Technologies and Internet of Things*, 2020, pp. 51–58.

[8] S. Dhelim, N. Aung, and H. Ning, "Mining user interest based on personality-aware hybrid filtering in social networks," *Knowledge-Based Syst.*, vol. 206, p. 106227, Oct. 2020.

[9] S. Dhelim, H. Ning, N. Aung, R. Huang, and J. Ma, "Personality-Aware Product Recommendation System Based on User Interests Mining and Metapath Discovery," *IEEE Trans. Comput. Soc. Syst.*, vol. 8, no. 1, pp. 86–98, Feb. 2021.

[10] S. Dhelim, H. Ning, M. A. Bouras, and J. Ma, "Cyber-Enabled Human-Centric Smart Home Architecture," in *2018 IEEE SmartWorld, Ubiquitous Intelligence & Computing, Advanced & Trusted Computing, Scalable Computing & Communications, Cloud & Big Data Computing, Internet of People and Smart City Innovations*, 2018, pp. 1880–1886.

[11] H. Ning, S. Dhelim, M. A. Bouras, A. Khelloufi, and A. Ullah, "Cyber-syndrome and its Formation, Classification, Recovery and Prevention," *IEEE Access*, 2018.

[12] S. Dhelim, N. Huansheng, S. Cui, M. Jianhua, R. Huang, and K. I.-K. Wang, "Cyberentity and its consistency in the cyber-physical-social-thinking hyperspace," *Comput. Electr. Eng.*, vol. 81, p. 106506, Jan. 2020.

[13] S. Dhelim, H. Ning, and N. Aung, "ComPath: User Interest Mining in Heterogeneous Signed Social Networks for Internet of People," *IEEE Internet Things J.*, pp. 1–1, 2020.

[14] F. Xia, N. Y. Asabere, H. Liu, Z. Chen, and W. Wang, "Socially Aware Conference Participant Recommendation With Personality Traits," *IEEE Syst. J.*, vol. 11, no. 4, pp. 2255–2266, Dec. 2017.

[15] M. F. Uddin, S. Banerjee, and J. Lee, "Recommender System Framework for Academic Choices: Personality Based Recommendation Engine (PBRE)," in *2016 IEEE 17th International Conference on Information Reuse and Integration (IRI)*, 2016, pp. 476–483.

[16] aAdli I. Hariadi and D. Nurjanah, "Hybrid attribute and personality based recommender system for book recommendation," in *2017 International Conference on Data and Software Engineering (ICoDSE)*, 2017, pp. 1–5.

[17] V. Zeigler-Hill and S. Monica, "The HEXACO model of personality and video game preferences," *Entertain. Comput.*, vol. 11, pp. 21–26, Nov. 2015.

[18] L. R. Goldberg, "An alternative" description of personality": the big-five factor structure.," *J. Pers. Soc. Psychol.*, vol. 59, no. 6, p. 1216, 1990.

[19] W. Revelle, "Hans Eysenck: Personality theorist," *Pers. Individ. Dif.*, vol. 103, pp. 32–39, 2016.

[20] M. C. Ashton and K. Lee, "Empirical, theoretical, and practical advantages of the HEXACO model of personality structure," *Personal. Soc. Psychol. Rev.*, vol. 11, no. 2, pp. 150–166, 2007.

[21] G. J. Boyle, "Myers-Briggs type indicator (MBTI): some psychometric limitations," *Aust. Psychol.*, vol. 30, no. 1, pp. 71–74, 1995.

[22] P. T. Costa Jr and R. R. McCrae, *The Revised NEO Personality Inventory (NEO-PI-R).* Sage Publications, Inc, 2008.

[23] O. P. John, E. M. Donahue, and R. L. Kentle, "Big five inventory," *J. Pers. Soc. Psychol.*, 1991.

[24] B. Rammstedt and O. P. John, "Measuring personality in one minute or less: A 10-item short version of the Big Five Inventory in English and German," *J. Res. Pers.*, vol. 41, no. 1, pp. 203–212, 2007.

[25] Y. Mehta, N. Majumder, A. Gelbukh, and E. Cambria, "Recent trends in deep learning based personality detection," *Artif. Intell. Rev.*, Oct. 2019.

[26] B. Lika, K. Kolomvatsos, and S. Hadjiefthymiades, "Facing the cold start problem in recommender systems," *Expert Syst. Appl.*, vol. 41, no. 4, pp. 2065–2073, 2014.